\documentclass[12pt]{iopart}
\usepackage{iopams}  
\usepackage{graphicx}
\usepackage{color}
\usepackage{cite}
\eqnobysec

\def\keywords#1{\vspace{10pt}
     \begin{indented}
     \item[]\rm Keywords: #1\par
     \end{indented}}

\def\be{\begin{equation}}
\def\ee{\end{equation}}
\def\bea{\begin{eqnarray}}
\def\eea{\end{eqnarray}}
\def\CF{{\mathcal F}}
\def\CG{{\mathcal G}}
\def\CH{{\mathcal H}}
\def\CN{{\mathcal N}}

\def\CZ{{\mathcal Z}}
\def\kBT{{k_{\mathrm B}T}}
\newcommand{\tens}[1]{\boldsymbol{\mathsf{#1}}}

\begin{document}
\jl{1}

\title[Ising model with multispin interactions]{One-dimensional Ising model with multispin interactions}

\author{Lo\"\i c Turban}

\address{Groupe de Physique Statistique, D\'epartement P2M, Institut Jean Lamour,\\ Universit\'e de Lorraine, CNRS (UMR 7198), 
Vand\oe uvre l\`es Nancy Cedex, F-54506, France} 

\ead{loic.turban@univ-lorraine.fr}

\begin{abstract}
We study the spin-$1/2$ Ising chain with multispin interactions $K$ involving the product of $m$ successive spins, for general values of $m$. Using a change of spin variables the zero-field partition function of a finite chain is obtained for free  and periodic boundary conditions (BC) and we calculate the two-spin correlation function. When placed in an external field $H$ the system is shown to be self-dual. Using another change of spin variables the one-dimensional (1D) Ising model with multispin interactions in a field is mapped onto a zero-field rectangular Ising model with first-neighbour interactions $K$ and $H$. The 2D system, with size $m\times N/m$, has the topology of a cylinder with helical BC. In the thermodynamic limit $N/m\to\infty$, $m\to\infty$, a 2D critical singularity develops on the self-duality line, $\sinh 2K\sinh 2H=1$.   
\end{abstract}


\submitto{\JPA}

\section{Introduction} 
The study of Ising models with $m$-spin interactions ($m>2$) have been an active field of research since the beginning of the seventies. The square-lattice eight-vertex model solved by Baxter~\cite{baxter71} was mapped onto an Ising model with two- and four-spin interactions by Wu~\cite{wu71} and, independently, by Kadanoff and Wegner~\cite{kadanoff71}. The Ashkin-Teller model~\cite{ashkin43}, a four-component system generalising the standard 2D Ising model, was also formulated as an Ising model on the square lattice with two- and four-spin interactions by Fan~\cite{fan72}. An Ising model with three-spin interactions on the triangular lattice was solved by Baxter and Wu~\cite{baxter73,baxter74}. One may also mention the pseudo-3D anisotropic Ising models with four-spin interactions solved by Suzuki~\cite{suzuki72}. Let us note that 2D and 3D Ising models in a field with multispin interactions of various forms have been studied through intensive Monte Carlo simulations~\cite{heringa89}. Most of these systems 
with multispin interactions have interesting duality properties~\cite{wegner71,merlini72,gruber77}.

Besides these models, then mostly of theoretical interest, systems with multiple binary variables were considered to describe equilibrium polymerisation~\cite{jaric83} and protein folding~\cite{hansen98,bakk02}. Randomly frustrated $p$-spin Ising models have been introduced to mimic spin glass behaviour~\cite{derrida80,derrida81,gardner85,deoliveira98,gillin01}, the limit $p\to\infty$ corresponding to the exactly solvable random-energy model~\cite{derrida80,derrida81}.

In the present work we study the 1D Ising model with Hamiltonian 
\be\fl
-\beta\CH_N[\{\sigma\}]=K\sum_{k} \underbrace{\sigma_k\sigma_{k+1}\sigma_{k+2}\cdots\sigma_{k+m-1}}_{\mbox{$m$ spins}}
+H\sum_{k}\sigma_k\,,
\qquad \beta=(\kBT)^{-1}\,.
\label{H}
\ee
The multispin interaction $K$ involves the product of $m$ adjacent Ising spins, $\sigma_k=\pm 1$. The system of $N$ spins is placed in a field $H$. Note that the factor $\beta$ has been absorbed in $K$ and $H$.
The thermodynamic properties of the system follow from the partition function,
\be
\CZ_N=\Tr_{\{\sigma\}}\exp\left(-\beta\CH_N[\{\sigma\}]\right)\,,
\label{ZN}
\ee
where the trace, $\Tr_{\{\sigma\}}(\ldots)$, denotes a sum over the spin configurations, $\prod_k\sum_{\sigma_k=\pm 1}(\ldots)$.

At $H=0$, ground-state configurations are obtained by the periodic repetition of the same pattern of $m$ spins, leading to a spin product $\prod_{l=0}^{m-1}\sigma_{k+l}$ equal to +1 (-1) for $K>0$ ($K<0$). There are $2^{m-1}$ ways to construct these degenerate ground-states~\cite{turban82a}. For example when $m=3$, at $H=0$ and $K>0$, the four degenerate ground-states are generated with the following patterns: $+++$, $+--$, $--+$, $-+-$.

The zero-field problem for any value of $m$ has been solved in the thermodynamic limit using a mapping onto the $k$-SAT problem on a ring~\cite{fan11}. A detailed solution for $m=3$ and $H\neq 0$ has been given in reference~\cite{mattis83}.
A 2D generalisation of~\eref{H} is obtained by coupling neighbouring 1D chains with multispin interactions through 
two-spin terms~\cite{turban82a}. In the strongly anisotropic limit, when the inter-chain coupling $K_\tau$ goes to infinity while the multispin interaction $K$ vanishes as $1/K_\tau$, the transfer operator at $H=0$ is related to a 1D quantum Ising chain with multispin interactions in a transverse field~\cite{turban82b,penson82}.

The paper is organised as follows: we consider a finite system for any value of $m$, at first in zero external field.  Using a change of spin variables we obtain exact expressions of the partition function for free BC in section 2 and periodic BC in section 3. We calculate the two-spin correlation function in section 4. Introducing the external field $H$, the system is shown to be self-dual in section 5. Using a new change of spin variables, the 1D Ising chain with multispin interactions $K$ in an external field $H$ is mapped onto a 2D rectangular Ising model with first-neighbour interactions $K$ and $H$. We conclude in section 7 and some calculations are detailed in two appendices.

%%%%%%%%%% FIG 1  %%%%%%%%%%%%%%%%%%%%%%%%%%%%%%%
\begin{figure}[t!]
\begin{center}
\includegraphics[width=10cm,angle=0]{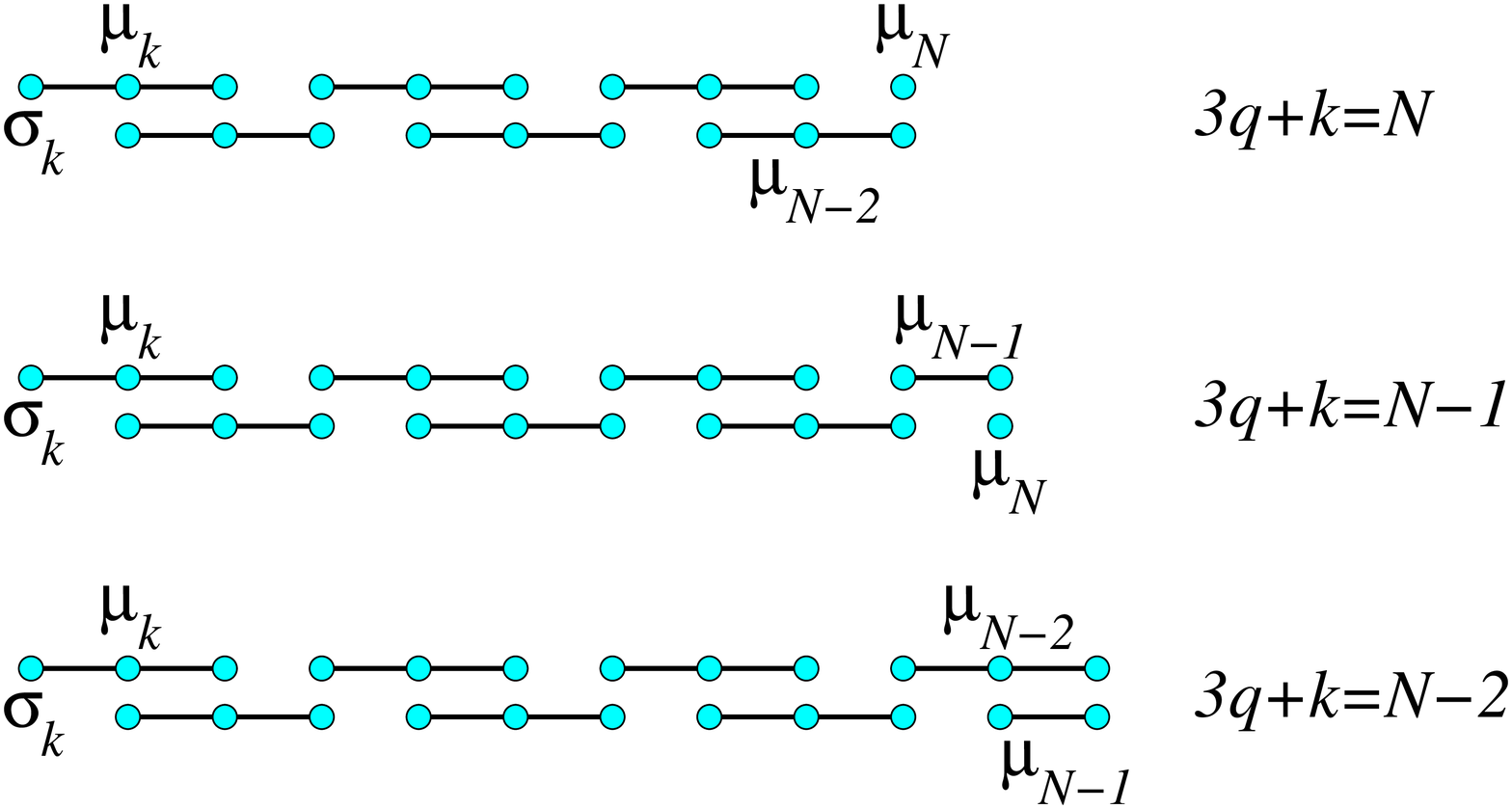}
\end{center}
\vglue -.0cm
\caption{Sets of $\mu$-variables entering into the expression of $\sigma_k$ \protect\eref{sigmak} when $m=3$, for different values of the distance $N-k$ to the end of the chain. The $\mu$-variables are built of triplets of $\sigma$-variables (circles) except for the two last ones, $\mu_{N-1}=\sigma_{N-1}\sigma_N$ and $\mu_N=\sigma_N$. All the $\sigma$s except  $\sigma_k$ appear twice in the product in~\protect\eref{sigmak} and thus contribute a factor of 1.
\label{fig1}  
}
\end{figure}
%%%%%%%%%%%%%%%%%%%%%%%%%%%%%%%%%%%%%%%%%

\section{Partition function at \boldmath{$H=0$} for free BC}
For the zero-field problem with free BC the Hamiltonian reduces to:
\be
-\beta\CH_N^{(f)}[\{\sigma\}]=K\sum_{k=1}^{N-m+1} \prod_{l=0}^{m-1}\sigma_{k+l}\,.
\label{Hf}
\ee
The form of the interactions suggests the following change of Ising variables~\cite{turban93,garrod95,mueller16}:
\be
\mu_k=\prod_{l=0}^{m-1}\sigma_{k+l}\,,\qquad k=1,\ldots,N\,.
\label{muk}
\ee
For $k>N-m+1$ the expression of $\mu_k$ involves non-existing spins $\sigma_k+l$ with $k+l>N$. The value of these
ghost spins is fixed to 1 so that they do not contribute to the product. Thus we have:
\be\fl
\mu_{N\!-m+2}\!=\!\underbrace{\sigma_{N\!-m+2}\sigma_{N\!-m+3}\cdots\sigma_N}_{\mbox{$m-1$ spins}}\,,\quad
\mu_{N\!-m+3}\!=\!\underbrace{\sigma_{N\!-m+3}\sigma_{N\!-m+4}\cdots\sigma_N}_{\mbox{$m-2$ spins}}\,,\ldots\,,
\mu_N\!=\!\sigma_N\,.
\label{muend}
\ee
There is a one-to-one relationship between old and new spin variables. The inverse transformation is given by:
\be\fl
\sigma_k=\prod_{r=0}^q\mu_{mr+k}\,\mu_{mr+k+1}\,,\qquad mq+k=N-l\,,\qquad l=0,\ldots,m-1\,.
\label{sigmak}
\ee
The $\sigma$s are non-local when expressed with the $\mu$s.
See figure~\ref{fig1} for an illustration of this relations for $m=3$~\footnote[1]{Note that for $l=0$ the last spin in~\protect\eref{sigmak} is $\mu_{N+1}=1$ since it is a product $m$ ghost spins. Thus, the last contribution to the product then comes from $\mu_N$ alone as shown in~figure~\protect\ref{fig1}.}.
Any set of $\sigma$s lead to a unique set of $\mu$s  and vice versa. $\mu_N$ and $\sigma_N$ are equal, 
the values of $\mu_{N-1}$ and $\sigma_{N-1}$ are related, and so on.

Using~\eref{muk} the Hamiltonian in~\eref{Hf} takes the following form:
\be
-\beta\CH_N^{(f)}[\{\mu\}]=K\sum_{k=1}^{N-m+1} \mu_k\,.
\label{Hfmu}
\ee
Note that a field term $H\sum_k\sigma_k$ in~\eref{Hf} would transform into a sum of highly non-local interactions 
involving the strings of $\mu$-variables in~\eref{sigmak}.
The new spin variables are non-interacting and the value of $m$ enters only through the number of spins.
Thus, for free BC, the partition function is given by:
\bea
\CZ_N^{(f)}&=\Tr_{\{\mu\}}\exp\left(-\beta\CH_N^{(f)}[\{\mu\}]\right)=
\underbrace{\prod_{k=1}^{N-m+1}\!\!\Tr_{\mu_k}\e^{K\mu_k}}_{(2\cosh K)^{N-m+1}}
\,\underbrace{\prod_{l=N-m+2}^{N}\!\!\!\!\Tr_{\mu_l}1}_{2^{m-1}}\nonumber\\
&=2^N(\cosh K)^{N-m+1}\,.
\label{ZNf}
\eea
The free energy can be written as
\be
\CF_N^{(f)}=-\kBT\ln\CZ_N^{(f)}=Nf_b+\CF_s(m)\,,
\label{FNf}
\ee
where
\be
f_b=-\kBT\ln(2\cosh K)\,,
\label{fb}
\ee
is the bulk free energy per spin, in agreement with reference~\cite{fan11}. It is independent of $m$, whereas the surface free energy
\be
\CF_s(m)=(m-1)\kBT\ln(\cosh K)\,,
\label{Fsm}
\ee
do depend on $m$.

\section{Partition function at \boldmath{$H=0$} for periodic BC}

For periodic BC we consider a system with a number of spins which is a multiple of $m$, $N=mp$, in order to respect the periodicity of the degenerate ground states. The Hamiltonian then takes the following form:
\be
-\beta\CH_N^{(p)}[\{\sigma\}]=K\sum_{k=1}^{N=mp}\,\prod_{l=0}^{m-1}\sigma_{k+l}\,,\qquad \sigma_{N+k}\equiv\sigma_k\,.
\label{Hp}
\ee
Using the $\mu$-variables defined in~\eref{muk}, together with the BC, $\sigma_{N+k}\equiv\sigma_k$, it can be rewritten as:
\be
-\beta\CH_N^{(p)}[\{\mu\}]=K\sum_{k=1}^{N=mp} \mu_k\,,
\label{Hpmu}
\ee
But due to periodic BC~\cite{turban93,mueller16}: 
\begin{itemize}
\item The correspondence between old and new variables is no longer one-to-one. Actually $2^{m-1}$ different 
$\sigma$-configurations lead to the same $\mu$-configuration.
\item All the $\mu$-configurations are not allowed. The $\mu$-variables are no longer independent, they have to satisfy $m-1$ constraints.
\end{itemize}

\begin{figure}[t!]
\begin{center}
\includegraphics[width=10cm,angle=0]{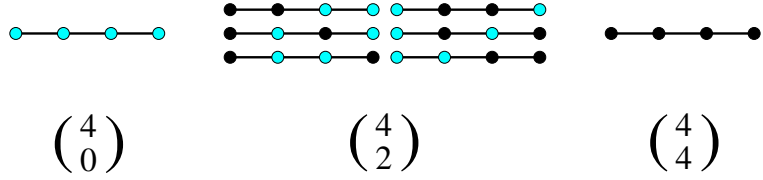}
\end{center}
\vglue -.0cm
\caption{The $2^3$ $\sigma$-configurations leading to the same $\mu$-variable when $m=4$. Circles correspond to $\sigma$ spins, black circles to flipped  $\sigma$ spins.
\label{fig2}  
}
\end{figure}

Consider the new spin variable $\mu_k=\prod_{l=0}^{m-1}\sigma_{k+l}$. It keeps the same value when an even number of the $\sigma$s are flipped (see figure~\ref{fig2}). Thus the number of configurations of the $\sigma$s leading to the same value of $\mu_k$ is given by
\be
g_m=\sum_{j=0}^{\lfloor m/2\rfloor}{m\choose 2j}=2^{m-1}\,,
\label{g}
\ee
where $j=0$ counts the initial configuration.
 
 Note that $\mu_k$ and $\mu_{k+1}$ have $m-1$ spins in common, $\prod_{l=1}^{m-1}\sigma_{k+l}$. When the number of flips for these common spins is odd, $\sigma_k$ and $\sigma_{k+m}$ have to be flipped in order to leave $\mu_k$ and $\mu_{k+1}$ unaffected. On the contrary, $\sigma_k$ and $\sigma_{k+m}$ must keep their original values when the number of common flips is even. The same is true for $\mu_{k+1}$ and $\mu_{k+2}$, and so on. It follows that the distribution of the flips is periodic  with period $m$~\footnote[2]{This is why $N$ has to be a multiple of $m$.}. Once the flips have been chosen for $\mu_k$, 
 there is no freedom left for the rest of the system. Thus $g_m=2^{m-1}$ gives the number of $\sigma$-configurations leading to the same $\mu$-configuration. Note that $g_m$ gives the ground-state degeneracy of $\CH_N[\{\sigma\}]$ when $H=0$.

%%%%%%%%%% FIG 2 %%%%%%%%%%%%%%%%%%%%%%%%%%%%%%%
%%%%%%%%%%%%%%%%%%%%%%%%%%%%%%%%%%%%%%%%%
 
\begin{figure}[t!]
\begin{center}
\includegraphics[width=11.14cm,angle=0]{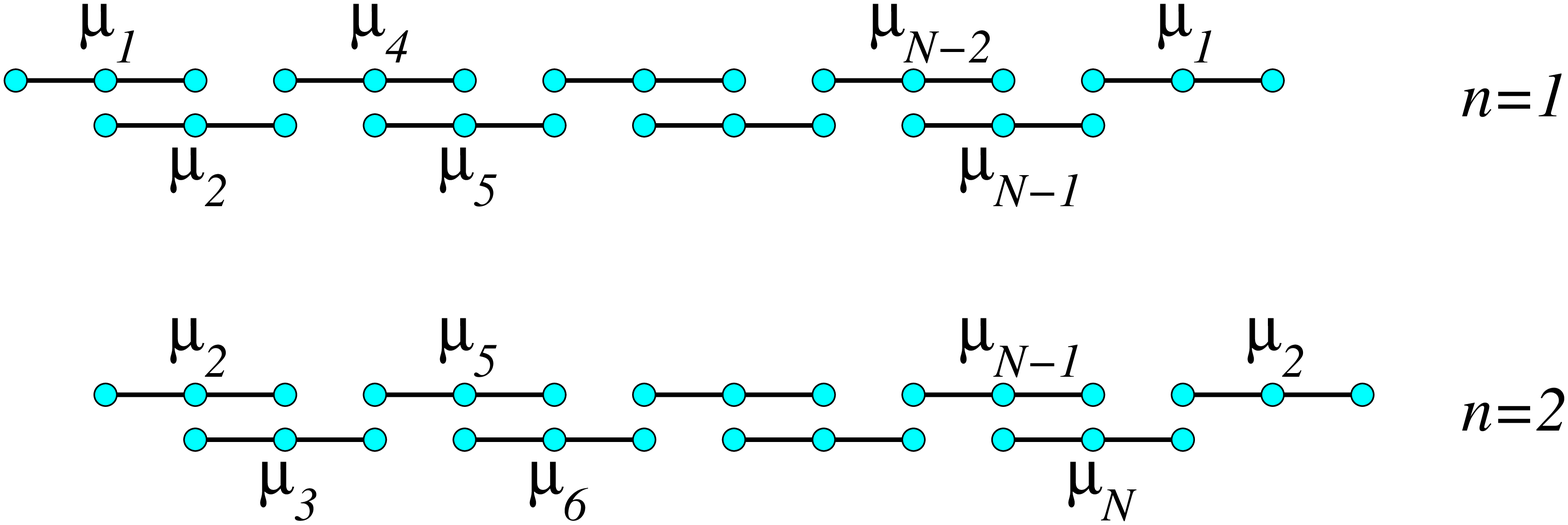}
\end{center}
\vglue -.0cm
\caption{Sets of $\mu$-variables entering into the expression of the constraints~\protect\eref{muprodq} for $m=3$ and $N=12$.
All the $\sigma$s (circles) appear in pairs so that the products are equal to 1.
\label{fig3}  
}
\end{figure}

 Let us now consider the constraints that the $\mu$ variables have to satisfy. Products of the form $\mu_{mq+n}\mu_{mq+n+1}$ with $n=1,\ldots,m-1$ can be rewritten as 
 \be
 \mu_{mq+n}\,\mu_{mq+n+1}=\sigma_{mq+n}\prod_{l=1}^{m-1}(\sigma_{mq+n+l})^2\sigma_{m(q+1)+n}=\sigma_{mq+n}\,\sigma_{m(q+1)+n}\,,
 \label{muprod}
 \ee
 using~\eref{muk}. In a product over $q$ from $0$ to $p-1$ all the remaining $\sigma$s appear twice thus the following set of constraints have to be imposed to the $\mu$-variables:
 \be
 \prod_{q=0}^{p-1}\mu_{mq+n}\,\mu_{mq+n+1}=1\,,\qquad n=1,\ldots,m-1\,.
 \label{muprodq}
 \ee
 See figure~\ref{fig3} for an illustration of these constraints when $m=3$. Note that other constraints can be defined but they follow from the fundamental ones given above. For instance, terms of the form $\mu_{mq+1}\,\mu_{mq+3}$ are obtained as the product of $\mu_{mq+1}\,\mu_{mq+2}$ by $\mu_{mq+2}\,\mu_{mq+3}$.

These constraints can be implemented using the Kronecker delta representation,
\be\fl
\delta_{P_n,1}=\case{1}{2}(1+P_n)\,,\qquad P_n=\prod_{q=0}^{p-1}\mu_{mq+n}\,\mu_{mq+n+1}=\pm 1\,,\qquad n=1,\ldots,m-1\,,
\label{Pn} 
\ee
to eliminate the states for which $P_n=-1$ in the partition sum over $\{\mu\}$. Taking into account the periodicity in the expressions of the constraints, it is convenient to write the Boltzmann factor as:
\be
\exp\left(-\beta\CH_N^{(p)}[\{\mu\}]\right)=\prod_{r=0}^{p-1}\prod_{l=1}^m\e^{K\mu_{mr+l}}\,.
\label{boltz}
\ee
Thus the partition function takes the following form,
\be
\CZ_{N=mp}^{(p)}=2^{m-1}\Tr_{\{\mu\}}\prod_{q=0}^{p-1}\prod_{l=1}^m\e^{K\mu_{mq+l}}
\prod_{n=1}^{m-1}\left(\frac{1+P_n}{2}\right)\,,
\label{ZNp}
\ee
where the first factor takes into account the multiplicity of the $\{\sigma\}$s for a given $\{\mu\}$ and the last product ensures the satisfaction of the $m-1$ constraints. The denominator in this last product cancels the front factor $2^{m-1}$
and we are left with the following expansion:
\be
\prod_{n=1}^{m-1}(1+P_n)=1+\sum_{n=1}^{m-1}P_n+\sum_{1\leq l<n\leq m-1}\!\!\!P_lP_n+\cdots+\prod_{n=1}^{m-1}P_n\,.
\label{exp}
\ee
Since the expression of $P_n$ in equation~\eref{Pn} is periodic with period $m$ one can examine the structure of the expansion on the first period ($q=0$) alone and restore the product over $q$ in each term afterwards. So let us consider the product
$\prod_{n=1}^{m-1}(1+\mu_n\mu_{n+1})$. Since $\mu_k^2=1$ the different spins appear in the expansion in pairs, quadruplets, etc. Actually, besides 1, the expansion generates products involving all the combinations of even numbers of different spins taken 
from the set $\{\mu_1,\mu_2,\ldots,\mu_m\}$. The number of terms obtained in this way is equal to $2^{m-1}-1$ as required (see~\eref{g}). For instance, with $m=4$, one obtains:
\bea\fl
&\mu_1\mu_2+\mu_2\mu_3+\mu_3\mu_4
+\mu_1\mu_2^2\mu_3+\mu_2\mu_3^2\mu_4+\mu_1\mu_2\mu_3\mu_4+\mu_1\mu_2^2\mu_3^2\mu_4\nonumber\\
&=\mu_1\mu_2+\mu_2\mu_3+\mu_3\mu_4+\mu_4\mu_1+\mu_1\mu_3+\mu_2\mu_4+\mu_1\mu_2\mu_3\mu_4\,.
\label{expm4}
\eea

%%%%%%%%%% FIG 3  %%%%%%%%%%%%%%%%%%%%%%%%%%%%%%%
%%%%%%%%%%%%%%%%%%%%%%%%%%%%%%%%%%%%%%%%%

The original expansion in~\eref{exp} can be rewritten as
\be
\prod_{n=1}^{m-1}(1+P_n)=1+\sum_{l=1}^{\lfloor m/2\rfloor}\sum_{\alpha_l=1}^{{m\choose 2l}}\prod_{q=0}^{p-1}\Xi_{\alpha_l}(q)\,,
\label{exp1}
\ee
where $\alpha_l$ denotes a combination of $2l$ spins taken from $m$ and
$\Xi_{\alpha_l}(q)$ is the product of $2l$ spins corresponding to the combination $\alpha_l$ for the $q$th cell. 

Going back to~\eref{ZNp}, one may write:
\be\fl
\CZ_{N=mp}^{(p)}=\Tr_{\{\mu\}}\prod_{q=0}^{p-1}\prod_{l=1}^m\e^{K\mu_{mq+l}}+\sum_{l=1}^{\lfloor m/2\rfloor}\sum_{\alpha_l=1}^{{m\choose 2l}}\Tr_{\{\mu\}}\prod_{q=0}^{p-1}\prod_{l=1}^m\e^{K\mu_{mq+l}}\Xi_{\alpha_l}(q)\,.
\label{ZNp1}
\ee
Summing over the spin configurations, the first term on the right gives $(2\cosh K)^{mp}$. In the second term, each combination of $2l$ spins, $\alpha_l$, contributes a factor of the form $2^m(\cosh K)^{m-2l}(\sinh K)^{2l}$ for each value of $q$ so that, finally:
\bea
\CZ_{N=mp}^{(p)}&=(2\cosh K)^{mp}+\sum_{l=1}^{\lfloor m/2\rfloor}{m\choose 2l}2^{mp}(\cosh K)^{(m-2l)p}(\sinh K)^{2lp}\nonumber\\
&=(2\cosh K)^N\left[1+\sum_{l=1}^{\lfloor m/2\rfloor}{m\choose 2l}(\tanh K)^{2lp}\right]\,.
\label{ZNp2}
\eea
Let $\tens{T}$ be the site-to-site transfer matrix. From the above expression of the partition function the eigenvalues of $\tens{T}^m$, $\omega_l=(2\cosh K)^m(\tanh K)^{2l}$ ($l=0,\lfloor m/2\rfloor$), and their degeneracy, $g_l={m\choose 2l}$, can be deduced (see appendix~A).

The free energy is given by
\be\fl
\CF_N^{(p)}=-\kBT\ln\CZ_{N=mp}^{(p)}=Nf_b-\kBT\ln\left[1+\sum_{l=1}^{\lfloor m/2\rfloor}{m\choose 2l}(\tanh K)^{2lN/m}\right]\,,
\label{FNp}
\ee
where the $m$-dependant finite-size correction to the bulk term vanishes in the thermodynamic limit.

\section{Correlation function at \boldmath{$H=0$}}
The two-spin correlation function $\CG_N(k,k')$ on a chain with free BC is given by:
\be
\CG_N(k,k')=\langle\sigma_k\,\sigma_{k'}\rangle=\frac{\Tr_ {\{\sigma\}}\exp\left(-\beta\CH_N[\{\sigma\}]\right)\sigma_k\,\sigma_{k'}}{\CZ_N}\,.
\label{GN}
\ee
For periodic BC a repeated use of~\eref{muprod} allows the expression of the correlation function using $\mu$ spins when the distance between the $\sigma$s is a multiple of $m$ but the implementation of the constraints~\eref{muprodq} is not easy. 
With free BC the inverse transformation~\eref{sigmak} can be used to re-express the correlation function for any value of the distance between the spins. 

Let us first consider the case where $k'-k$ is a multiple of $m$ so that $k=ms+n$, $k'=k+mt=m(s+t)+n$ with $n=0,\ldots,m-1$. Making use of~\eref{sigmak}, one obtains:
\be
\sigma_{ms+n}\,\sigma_{m(s+t)+n}=\prod_{r=s}^{s+t-1}\mu_{mr+n}\,\mu_{mr+n+1}\,.
\label{sigmasigma}
\ee
Then the numerator in~\eref{GN} can be rewritten as:
\be
\CN=\Tr_{\{\mu\}}\prod_{k=1}^{N-m+1}\e^{K\mu_k}\prod_{r=s}^{s+t-1}\mu_{mr+n}\,\mu_{mr+n+1}\,.
\label{num}
\ee
In this expression three different types of traces are involved: $\Tr_{\mu_k}\e^{K\mu_k}=2\cosh K$, $\Tr_{\mu_k}\e^{K\mu_k}\mu_k=2\sinh K$ and $\Tr_{\mu_k}1=2$. There are $N-m+1-2t$ factors of the first type, $2t$ factors of the second type and $m-1$ factors of the third type so that:
\be\fl
\CN=2^{m-1}(2\cosh K)^{N-m+1-2t}(2\sinh K)^{2t}=2^N(\cosh K)^{N-m+1}(\tanh K)^{2t}\,.
\label{num1}
\ee
Inserting this expression in~\eref{GN} and using~\eref{ZNf} leads to
\be
\CG_N(k,k+mt)=(\tanh K)^{2t}=\e^{-mt/\xi}\,,
\label{GN1}
\ee
where $\xi$ is the $m$-dependant correlation length given by:
\be
\xi=-\frac{m}{2\ln|\tanh K|}\,.
\label{xi}
\ee
Note that the correlation function in~\eref{GN1} is independent of $k$ because the $\mu$ spins are non-interacting.

%%%%%%%%%% FIG 4 %%%%%%%%%%%%%%%%%%%%%%%%%%%%%%%
\begin{figure}[t!]
\begin{center}
\includegraphics[width=10.37cm,angle=0]{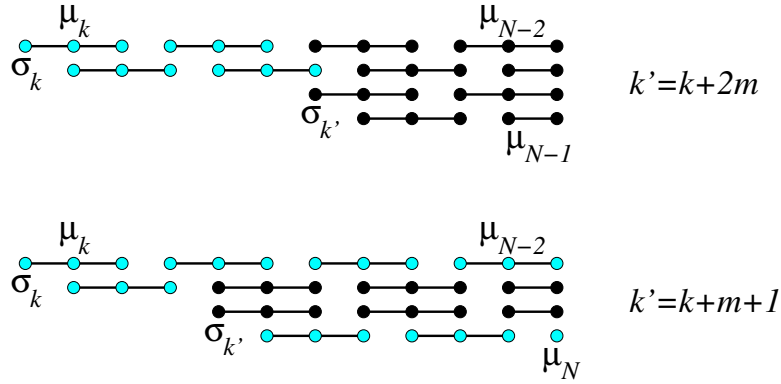}
\end{center}
\vglue -.0cm
\caption{When $k'-k$ is a multiple of $m$ (top) the product $\sigma_k\,\sigma_{k'}$ in~\protect\eref{sigmasigma} contains only $\mu_j$s with $k\leq j<k'$. All other $\mu$s (with black circles) appear in pairs. Otherwise (bottom), for $m>2$, $\sigma_k\,\sigma_{k'}$ involves a string of $\mu$s going on until the end of the chain with at least one unpaired $\mu_j$ with $j>N-m+1$ (here $j=N$) which makes the trace vanish. 
\label{fig4}  
}
\end{figure}
%%%%%%%%%%%%%%%%%%%%%%%%%%%%%%%%%%%%%%%%%

Suppose now that $k=ms+n$, $k'=m(s+t)+n'$ with $0\leq n,n'\leq m-1$ and $n\neq n'$. When $m>2$ the product $\sigma_k\,\sigma_{k'}$ involves a string of $\mu$s which does not stop before $k'$ since the remaining ones no longer appear systematically in pairs (see figure~\ref{fig4})~\footnote[3]{The case $m=2$ is special in that all the $\mu_j$s with $j\geq k$ enter into the 
expression~\protect\eref{sigmak} of $\sigma_k$.}. In particular there is at least one unpaired $\mu_j$ with $j>N-m+1$ making the trace  vanish. It follows that for $m>2$: 
\be
\CG_N(k,k+mt+n'')=0\,,\qquad n''=n'-n=1,\ldots,m-1\,.
\label{GN2}
\ee
This result can be recovered by taking into account the invariance of the Hamiltonian under the periodic flip of two spins for each period of $m$ spins, as discussed below~\eref{g}. Consider the correlation function 
$\langle\sigma_{ms+n}\,\sigma_{m(s+t)+n'}\rangle$ with $0\leq n,n'\leq m-1$ and $n\neq n'$. The trace over $\{\sigma\}$ is not affected by the periodic change of spin variables, 
\be\fl
\sigma_{mr+n'}\to-\sigma_{mr+n'}\,,\quad \sigma_{mr+n''}\to-\sigma_{mr+n''}\,,\quad 0\leq n''\leq m-1\,,\quad n''\neq n,n'\,,
\label{change}
\ee
so that, due the the change of sign in the spin product $\sigma_{ms+n}\,\sigma_{m(s+t)+n'}$, one obtains:
\be
\langle\sigma_{ms+n}\,\sigma_{m(s+t)+n'}\rangle=-\langle\sigma_{ms+n}\,\sigma_{m(s+t)+n'}\rangle=0\,.
\label{GN3}
\ee
Note that for $m=2$ this argument does not apply since there is no place left for $n''$ in the interval $[0,m-1]$ once $n$ and $n'$ have been chosen.

\section{Self-duality under external field}

%%%%%%%%%% FIG 5 %%%%%%%%%%%%%%%%%%%%%%%%%%%%%%%
\begin{figure}[t!]
\begin{center}
\includegraphics[width=9.3cm,angle=0]{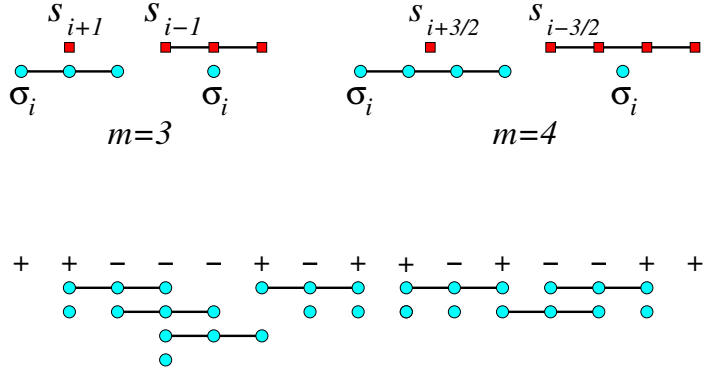}
\end{center}
\vglue -.0cm
\caption{Top: position of the dual spin variables relative to the initial interactions for $m=3$ and~$m=4$.
The two lattices coincide when $m$ is odd. The dual lattice is shifted by one-half of a lattice constant when $m$ is even. When a dual spin or a product of dual spins is negative, the corresponding initial interaction contributes to a non-vanishing term in the high-temperature expansion~\protect\eref{ZNKH1}. Bottom: a dual spin configuration and the corresponding non-vanishing high-temperature diagram for $m=3$ and periodic BC. The $\sigma$s under the dual spins systematically appear an even number of times on each site. 
\label{fig5}  
}
\end{figure}
%%%%%%%%%%%%%%%%%%%%%%%%%%%%%%%%%%%%%%%%%

The partition function for a non-vanishing external field $H$ and periodic BC is given by:
\be 
\CZ_N^{(p)}(K,H)=\Tr_{\{\sigma\}}\prod_{k=1}^N\exp\left(K\prod_{l=0}^{m-1}\sigma_{k+l}\right)\exp(H\sigma_k)\,.
\label{ZNKH}
\ee
Making use of the high-temperature expansion,
\bea
&\exp\left(K\prod_{l=0}^{m-1}\sigma_{k+l}\right)=\cosh K\sum_{u_k=0,1}\!\!\left(\tanh K\prod_{l=0}^{m-1}\sigma_{k+l}\right)^{u_k}\,,\nonumber\\
&\exp(H\sigma_k)=\cosh H\sum_{v_k=0,1}\!\!(\sigma_k\tanh H)^{v_k}\,,
\label{hTe}
\eea
and collecting the $\sigma_i$s attached to the same site~$i$, equation~\eref{ZNKH} can be rewritten as
\bea
\fl&\CZ_N^{(p)}(K,H)=(\cosh K\cosh H)^N\Tr_{\{u,v\}}\prod_{k=1}^N(\tanh K)^{u_k}(\tanh H)^{v_k}
\Tr_{\{\sigma\}}\prod_{i=1}^N\sigma_i^{\omega_i}\,,\nonumber\\
\fl&\ \ \ \ \ \ \ \ \ \ \ \ \omega_i=v_i+\sum_{l=0}^{m-1}u_{i-l}\,,
\label{ZNKH1}
\eea
where $\Tr_{\{u,v\}}$ is a sum over the new variables $u_k,v_k=0,1$. The trace over $\{\sigma\}$ is non-vanishing and leads to a factor $2^N$ only when all the $\omega_i$s are even. This can be systematically realised~\cite{savit80} by relating $u_i$ and $v_i$ to dual Ising spin variables $s_i=\pm1$ as follows (see figure~\ref{fig5}):
\be
u_i=\frac{1}{2}\left(1-s_{i+(m-1)/2}\right)\,,\qquad v_i=\frac{1}{2}\left(1-\prod_{l=0}^{m-1}s_{i+l-(m-1)/2}\right)\,.
\label{uvs}
\ee
Note that the original lattice and its dual coincide when $m$ is odd whereas the dual lattice is shifted by $1/2$ when $m$ is even~(see top of figure~\ref{fig5}). Let us consider the expression of the $\sigma$-exponent $\omega_i$ in terms of dual spins:
\be
\omega_i\!=\!\frac{m\!+\!1}{2}-\!\frac{1}{2}\left(\prod_{l=0}^{m-1}s_{i+l-(m-1)/2}\!+\!\sum_{l=0}^{m-1}s_{i-l+(m-1)/2}\right).
\label{exps}
\ee
There are $m$ dual spins entering in this expression from $s_{i-(m-1)/2}$ to $s_{i+(m-1)/2}$, each spin contributing once in the sum and in the product. A dual configuration with all these spins equal to $+1$ gives $\omega_i=0$. When one of these spins is flipped, two terms in the bracket change sign. This remains true with two flips because the product keeps its initial value. More generally, with $n$ flips on different dual spins, $2\times\lfloor(n+1)/2\rfloor$ terms in the bracket change sign. Hence any dual spin configuration leads to an even exponent $\omega_i$ and to a non-vanishing graph in the high-temperature expansion. Thus the trace over $\{u,v\}$ in~\eref{ZNKH1} can be replaced by a trace over the dual spins $\{s\}$~\footnote[4]{Note that, due to the single-spin terms in~\eref{uvs}, each dual spin configuration leads to a different diagram in the high-temperature expansion and vice-versa (see figure~\ref{fig5} for an illustration).}. 
The partition function takes the following form,  
\bea
\fl &\CZ_N^{(p)}(K,H)=(2\cosh K\cosh H)^N\Tr_{\{s\}}\prod_{k=1}^N(\tanh K)^{u_k}(\tanh H)^{v_k}\nonumber\\
\fl &\ \ \ \ \ \ \ \ \ =(\sinh 2K\sinh 2H)^{N/2}\Tr_{\{s\}}\exp\left(\widetilde{K}\sum_{k=1}^N\prod_{l=0}^{m-1}s_{k+l-(m-1)/2}
+\widetilde{H}\sum_{k=1}^{N}s_{k+(m-1)/2}\right)\nonumber\\
\fl &\ \ \ \ \ \ \ \ \ =(\sinh 2K\sinh 2H)^{N/2}\CZ_N^{(p)}(\widetilde{K},\widetilde{H})\,,
\label{dualZ}
\eea
where we introduced the dual couplings, 
\be
\widetilde{K}=-\frac{1}{2}\ln(\tanh H)\,,\qquad \widetilde{H}=-\frac{1}{2}\ln(\tanh K)\,,
\label{dualKH}
\ee
such that:
\be
\sinh 2\widetilde{K}\sinh 2H=1\,,\qquad \sinh 2\widetilde{H}\sinh 2K=1\,.
\label{dualKH1}
\ee
As a consequence,
\be
\sinh 2K\sinh 2H=\frac{1}{\sinh 2\widetilde{K}\sinh 2\widetilde{H}}\,,
\label{sd}
\ee
and
\be
\sinh 2K\sinh 2H=\pm1\,,
\label{sdl}
\ee
which is invariant in the transformation, is a self-duality line.
Equation~\eref{sd} can be used to rewrite~\eref{dualZ} under the symmetric form:
\be
\frac{\CZ_N^{(p)}(K,H)}{(\sinh 2K\sinh 2H)^{N/4}}=\frac{\CZ_N^{(p)}(\widetilde{K},\widetilde{H})}{(\sinh 2\widetilde{K}\sinh 2\widetilde{H})^{N/4}}\,.
\label{dualZ1}
\ee
The duality relations keep the same form for any value of $m$.

\section{Mapping on a 2D Ising model when \boldmath{$H\neq0$}}

%%%%%%%%%% FIG 6  %%%%%%%%%%%%%%%%%%%%%%%%%%%%%%%
\begin{figure}[t!]
\begin{center}
\includegraphics[width=8.4cm,angle=0]{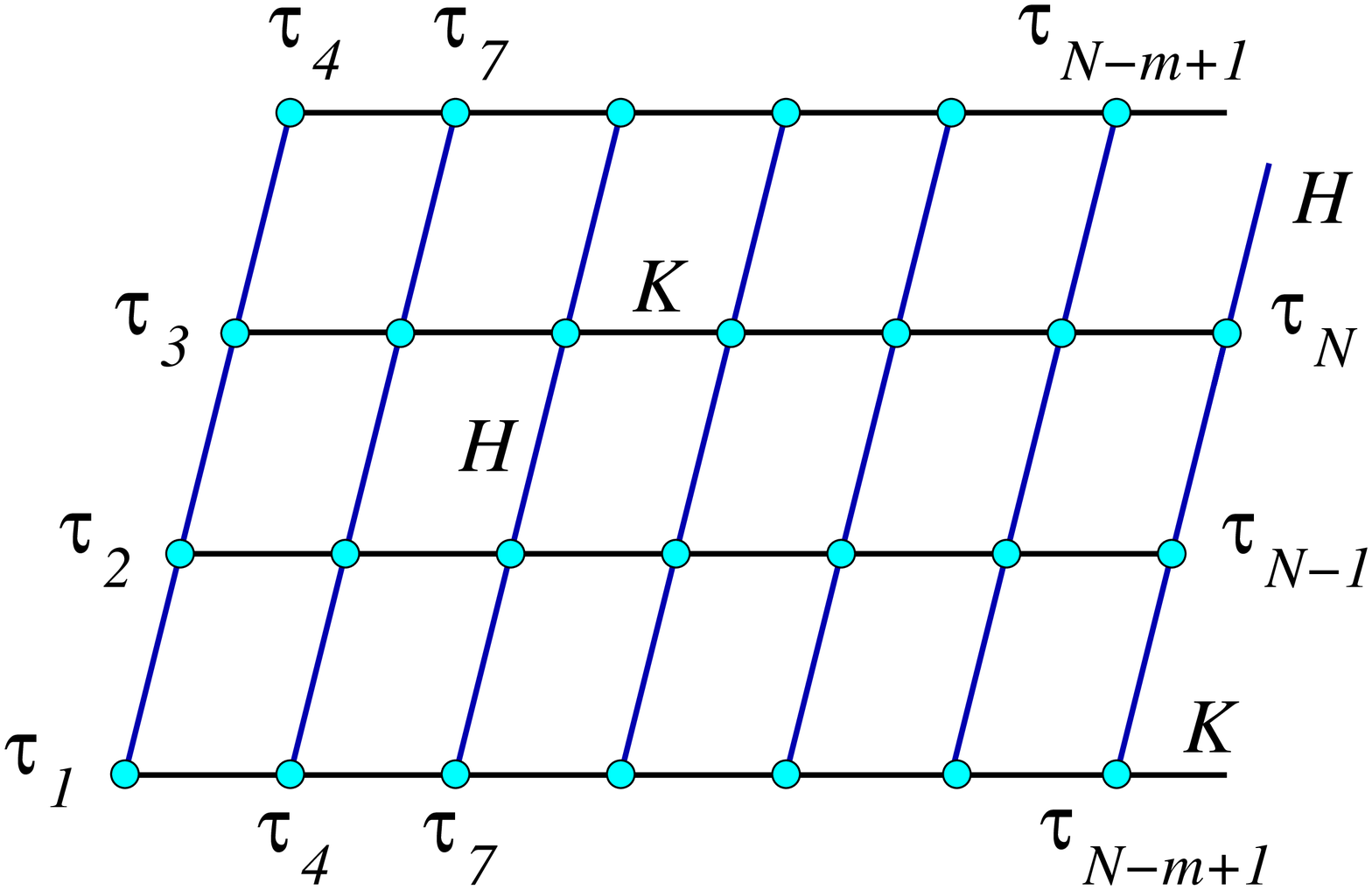}
\end{center}
\vglue -.0cm
\caption{Under the change of spin variables~\protect\eref{tauk}, the 1D Ising model with $m$-spin interactions in an external field is mapped onto a rectangular Ising model on a cylinder with helical BC. The first-neighbour interactions are equal to $H$ along the helix and $K$ parallel to the cylinder axis. There are $m$ spins per turn and $N/m$ turns (here, $m=3$ and $N=21$).
\label{fig6}  
}
\end{figure}
%%%%%%%%%%%%%%%%%%%%%%%%%%%%%%%%%%%%%%%%%

We now consider the multispin Ising model in a field $H$ with free BC and $N=mp$. The Hamiltonian reads:
\be
-\beta\CH_N^{(f)}[\{\sigma\}]=K\sum_{k=1}^{N-m+1} \prod_{l=0}^{m-1}\sigma_{k+l}+H\sum_{k=1}^N\sigma_k\,.
\label{Hf1}
\ee
Let us introduce a new set of Ising variables, $\{\tau\}$, given by~\cite{suzuki72}: 
\be
\tau_k=\prod_{i=k}^N\sigma_i\,.
\label{tauk}
\ee
Thus we have
\be
\sigma_k=\left\{
\begin{array}{ll}
\tau_{k}\,\tau_{k+1}& k=1,\ldots,N-1\,,\\
\ms\ms
\tau_N              & k=N\,,
\end{array}
\right.
\label{sigmatau}
\ee
so that there is a one-to-one correspondence between the two sets of spin variables. The multispin interaction in~\eref{Hf1} transforms as $\prod_{l=0}^{m-1}\sigma_{k+l}=\tau_k\,\tau_{k+m}$ except for $k=N-m+1$ where it gives $\tau_{N-m+1}$. For the interaction with the external field, $\sigma_k$ in~\eref{sigmatau} gives a first-neighbour interaction in the new variables except for the last spin. The transformed Hamiltonian takes the following form:
\be\fl
-\beta\CH_N[\{\tau\}]=K\sum_{k=1}^{N-m}\tau_k\,\tau_{k+m}+H\sum_{k=1}^{N-1}\tau_k\,\tau_{k+1}+K\tau_{N-m+1}+H\tau_N\,,
\qquad N=mp\,.
\label{Htau}
\ee
Thus we obtain a 2D Ising model on a cylinder with helical BC. The longitudinal size of the lattice is $N/m$ and there are $m$ spins per turn. The system is anisotropic with two-spin interactions $K$ in the direction of the cylinder axis and $H$ along the helix. Local fields, either $H$ or $K$, are acting on two of the end spins (see figure~\ref{fig6}). 

When $H=0$ one obtains $m$ non-interacting Ising chains with first-neighbour interactions $K$. For general values of $N=mp+l$ there are $l$ chains with $p+1$ spins, one chain with $p$ spins for which a field $K$ is acting on the last spin and $m-l-1$ chains with $p$ spins. The multiplicative contributions to the partition function for each chain are respectively $2^{p+1}(\cosh K)^p$, $2^p(\cosh K)^p$  and $2^p(\cosh K)^{p-1}$. Collecting these factors, the partition function $\CZ_N^{(f)}$ in~\eref{ZNf} is recovered.

With the new variables the spin-spin correlation function translates into
\be
\CG_N(k,k')=\langle\sigma_k\,\sigma_{k'}\rangle=\langle\tau_k\,\tau_{k+1}\,\tau_{k'}\,\tau_{k'+1}\rangle\,,
\label{corr2d}
\ee
i.e., into an energy-energy correlation function on the 2D lattice (see figure~\ref{fig6}). When $H=0$ and $k'=k+mt$ equation~\eref{corr2d} gives a product of correlation functions for two $\tau$-spins at a distance $t$ on two non-interacting Ising chains,
\be
\CG_N(k,k+mt)=\langle\tau_k\,\tau_{k+mt}\rangle\langle\tau_{k+1}\,\tau_{k+mt+1}\rangle\,,
\label{corrtau}
\ee
each of which contributes a factor $(\tanh K)^t$, in agreement with~\eref{GN1}. When $k'=k+mt+n$ with $n=1,\ldots,m-1$ there remains at 
least one unpaired $\tau$ with a vanishing average thus leading to~\eref{GN2}.

When $m=1$ the system corresponds either to $N$ non-interacting $\sigma$-spins in a field $K+H$ or a chain of $N$ $\tau$-spins with first-neigbour interactions $K+H$ and a field $K+H$ acting on the last spin. Using the high-temperature expansion, it is easy to verify that the partition function is then $[2\cosh(K+H)]^N$ in agreement with~\eref{ZNf} and~\eref{ZNp2} when $H=0$.

\begin{figure}[t!]
\begin{center}
\includegraphics[width=11.5cm,angle=0]{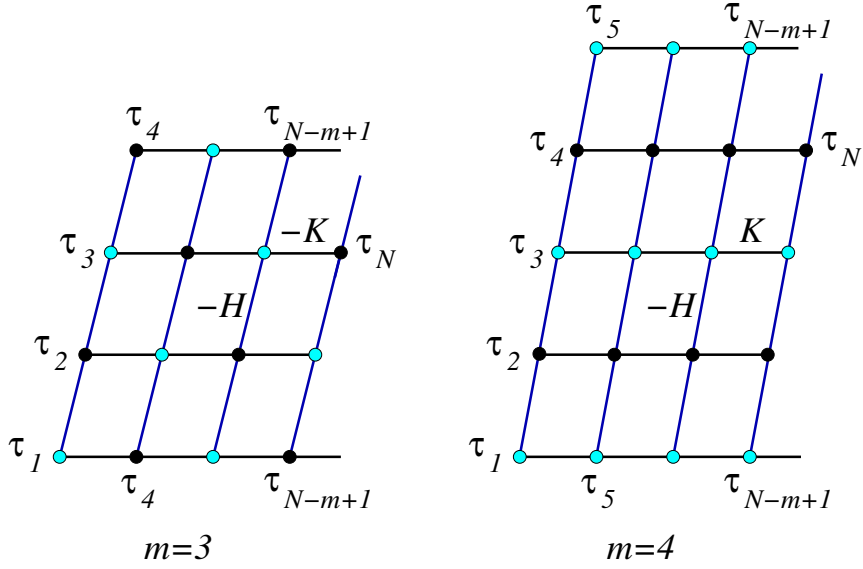}
\end{center}
\vglue -.0cm
\caption{Location of the spin flips (black circles) and new values of the couplings on the cylinder with helical boundary conditions resulting from the change of spin variables $\tau_k\to\tau'_k=(-1)^{N-k+1}\tau_k$ ($N=12$ for  $m=3$ and $N=16$ for $m=4$).
\label{fig7}  
}
\end{figure}

The symmetries of the partition function depends on the parity of $m$. Let us first examine the original 1D Ising chain. A change of spin variables $\{\sigma\}\to\{\sigma'=-\sigma\}$ does not affect the trace operation but modifies the Hamiltonian for which $H\to-H$ and $K\to(-1)^mK$ so that:
\be
\CZ_N(K,H)=\left\{
\begin{array}{ll}
\CZ_N(-K,-H)\,,\qquad& m\ \mathrm{odd}\,,\\
\ms\ms
\CZ_N(K,-H)\,,& m\ \mathrm{even}\,.
\end{array}
\right.
\label{symZ}
\ee
For the 2D system, according to~\eref{tauk}, $\tau_k\to\tau'_k=(-1)^{N-k+1}\tau_k$. Then for odd values of $m$ the spin flips are out of phase after one turn and both $K$ and $H$ change sign, whereas for even values of $m$, the spin flips stay in phase and $H$ alone changes sign, in agreement with~\eref{symZ}. This is illustrated  in figure~\ref{fig7} for $m=3$ and $m=4$. 

In the thermodynamic limit, $N/m\to\infty$, the free energy of the multispin chain in a field develops a 2D Ising critical singularity~\footnote[5]{The local external fields on the end spins do not affect the bulk behaviour.} on the self-duality line, $\sinh 2K\sinh 2H=1$ when $m\to\infty$. The free energy per spin is then given by the Onsager expression~\cite{onsager44}:
\be\fl
f_b\!=\!-\kBT\left[\ln2\!+\!\frac{1}{2\pi^2}\!\!\int_0^\pi\!\!\!\! d\theta\!\int_0^\pi\!\!\!\! d\varphi\ln(\cosh2K\cosh2H\!-\!\sinh2K\cos\theta\!
-\!\sinh2H\cos\varphi)\right]\!.
\label{fbKH}
\ee
The case of finite $m$ values is discussed in appendix~B.

%%%%%%%%%% FIG 7  %%%%%%%%%%%%%%%%%%%%%%%%%%%%%%%
%%%%%%%%%%%%%%%%%%%%%%%%%%%%%%%%%%%%%%%%%

\section{Conclusion}

At $H=0$, we have obtained exact results for the partition function of the finite-size 1D
Ising model with $m$-spin interactions $K$, for free  and periodic BC. The two-spin correlation function have been calculated for free BC. The eigenvalues of the $m$th power of the transfer matrix $\tens{T}$ and their degeneracy have been deduced from the expression of the partition function with periodic BC.
 
At $H\neq0$, the system with periodic BC is self-dual on the line $\sinh 2K\,\sinh 2H=1$. Our main result, obtained via a change of spin variables for a system with free BC, is a mapping of the 1D Ising model with multispin interactions $K$ in a field $H$ onto an anisotropic finite-size 2D Ising model with first-neighbour interactions $K$ and $H$. The 2D system, with size $m\times N/m$, has the topology of a cylinder with helical BC. Note that a change of spin variables similar to~\eref{tauk} have been used in the reverse direction by Suzuki~\cite{suzuki72} to map a 3D system with four-spin interactions onto a 2D system with two-spin interactions.

In the thermodynamic limit, $N\to\infty$, the free energy per spin of the multispin Ising model in a field can be deduced from known results for the 2D model. When $m\to\infty$ it is given by Onsager result~\cite{onsager44}. In this limit the 1D system is critical on the self-duality line. The self-duality of the 1D system in a field actually appears as a translation of the self-duality of the 2D system without external field. When $m$ is finite, the free energy per spin can be extracted from known results obtained by Liaw {\it et al}~\cite{liaw06} for the anisotropic 2D Ising model on a torus with helical boundary conditions in the limit of an infinite major radius.

Since a field derivative in 1D corresponds to a derivative with respect to a two-spin interaction in 2D, the two-spin correlation function in 1D becomes an energy-energy correlation function in 2D. The magnetisation and the susceptibility of the 1D multispin Ising model in a field have the same behaviour as the internal energy and the specific heat, respectively. Both display the 2D critical behaviour when $m\to\infty$. This correspondence is also a consequence of the $H-K$ duality. 

\ack
I would like to thank Christophe Chatelain and Jean-Yves Fortin for useful discussions.

\appendix

\section{Transfer matrix at \boldmath{$H=0$}}
\setcounter{section}{1}

When written in the basis
$\{|++\rangle,|+-\rangle,|-+\rangle,|--\rangle\}$,
the transfer matrix of the model with 3-spin interactions, from $|\sigma_k\sigma_{k+1}\rangle$ to $|\sigma_{k+1}\sigma_{k+2}\rangle$, takes the form
\be
\tens{T}=\left(
\begin{array}{cccc}
\e^K  &  \e^{-K}  &  0  &  0\\
  0   &   0  &  \e^{-K} &  \e^K\\
 \e^{-K} &  \e^K  &   0   &   0\\
  0  &  0  &  \e^K  &  \e^{-K}
\end{array}
\right)
\label{T}
\ee 
in a vanishing external field.
It is asymmetric and its eigenvalues are complex:
\be\fl
\lambda_1=2\cosh K\,,\qquad \lambda_{2,3,4}=2\left[\cosh K\,(\sinh K)^2\right]^{1/3}\e^{i k2\pi/3}\,,\quad k=0,1,2\,.
\label{eig}
\ee
This oscillating behaviour is linked to the periodicity of the degenerate ground-states. Taking the cube of the transfer matrix, i. e., transferring by one period from $k$ to $k+3$, one obtains a symmetric matrix 
\be
\tens{T}^3=\left(
\begin{array}{cccc}
2\cosh 3K & 2\cosh K  & 2\cosh K  &  2\cosh K \\
2\cosh K  & 2\cosh 3K & 2\cosh K  &  2\cosh K \\
2\cosh K  & 2\cosh K  & 2\cosh 3K &  2\cosh K \\
2\cosh K  & 2\cosh K  & 2\cosh K  &  2\cosh 3K
\end{array}
\right)\,,
\label{T3}
\ee 
with real eigenvalues: $8\cosh^3 K$ and $8\cosh K\,\sinh^2 K$, which is 3 times degenerate.

For any value of $m$ the $2^{m-1}$ eigenvalues of $\tens{T}^m$, $\omega_l$, and their degeneracy, $g_l$, can be extracted from 
the expression of the partition function with periodic BC. Since
\be
\CZ_{N=mp}^{(p)}=\Tr(\tens{T}^{m})^p=\sum_{l=0}^{\lfloor m/2\rfloor}g_l\,\omega_l^p
\label{ZNp3}
\ee
it follows from~\eref{ZNp2} that:
\be
\omega_l=(2\cosh K)^m(\tanh K)^{2l}\,,\qquad g_l={m\choose 2l}\,,\qquad l=0,\lfloor m/2\rfloor\,.
\label{eigm}
\ee

\section{Free energy per site in the thermodynamic limit when \boldmath{$H\neq 0$}}

The partition function of the rectangular Ising model with helical boundary conditions has been obtained on a torus in reference~\cite{liaw06}~\footnote[6]{A detailed study of finite-size effects in this geometry can be found in reference~\cite{izmailian07}, unfortunately for isotropic interactions, $K=H$.}.  
This exact result can be exploited in the limit where the major radius of the torus become infinite to calculate the free energy per site of our system with transverse size $m$ when $N\to\infty$. It corresponds to the thermodynamic limit of the Ising chain with multispin interactions in a field. Note that, in this limit, the local field terms on the end spins becomes irrelevant.

With our notations the partition function of the rectangular Ising model on the torus with size $L\times m$ ($L=N/m$), first-neighbour interactions $K$ and $H$, and  twisting factor $1/m$,  is given by~\cite{liaw06}
\bea
\fl&\CZ_{m,L}=(2\cosh K\,\cosh H)^{mL}Q_{m,L}\,,\nonumber\\
\fl&Q_{m,L}=\frac{1}{2}\left[I_{m,L}\left(\frac{1}{2},\frac{1}{2}\right)\!+\!I_{m,L}\left(\frac{1}{2},0\right)
\!+\!I_{m,L}\left(0,\frac{1}{2}\right)\!-\!\mathrm{sgn}\left(\frac{T\!-\!T_c}{T_c}\right)I_{m,L}\left(0,0\right)\right]\,,
\label{ZmL}
\eea
where $T_c$ is the critical temperature of the bulk system and
\bea
\fl&I_{m,L}(\alpha,\beta)=\prod_{p=1}^m\prod_{q=1}^L\left\{\lambda_0
-\lambda_1\cos\left[2\pi\left(\frac{p+\alpha}{m}-\frac{q+\beta}{mL}\right)\right]
-\lambda_2\cos\left[2\pi\left(\frac{q+\beta}{L}\right)\right]\right\}^{1/2}\,,\nonumber\\
\fl&\lambda_0=\frac{\cosh 2K\cosh 2H}{\cosh^2K\cosh^2H}\,,\qquad\lambda_1=2\,\frac{\tanh H}{\cosh^2 K}\,,\qquad
\lambda_2=2\,\frac{\tanh K}{\cosh^2 H}\,.
\label{ImL}
\eea
When $L\to\infty$, $I_{m,L}(\alpha,\beta)=I_{m,L}(\alpha,0)$ so that, in this limit, one may write:
\be
Q_{m,L}=\left\{
\begin{array}{ll}
I_{m,L}\left(\frac{1}{2},0\right)\,,& T>T_c\,,\\
\ms\ms
I_{m,L}\left(\frac{1}{2},0\right)+I_{m,L}(0,0)\,,& T<T_c\,.
\end{array}
\right.
\label{ImL1}
\ee
With
\be\fl
A_m(\alpha)=\lim_{L\to\infty}\frac{1}{mL}\ln\left[(\cosh K\,\cosh H)^{mL}I_{m,L}(\alpha,0)\right]\,,\qquad \theta=\lim_{L\to\infty}2\pi \frac{q}{L}\,,
\label{Am}
\ee
one obtains:
\bea
A_m(\alpha)=\frac{1}{4\pi m}\sum_{p=1}^m\int_0^{2\pi}\!\!&d\theta\ln\Biggl\{\cosh2K\cosh2H
-\sinh2K\cos\theta\Biggr.\nonumber\\
&\ \ \ \ \ \ \ \ \ \ \ \ -\left.\sinh2H\cos\left[\frac{\theta-2\pi(p+\alpha)}{m}\right]\right\}\,.
\label{Am1}
\eea
Thus, in the thermodynamic limit of the 1D Ising model with $m$-spin interaction $K$ in a field $H$, the free energy 
per spin is given by:
\be\fl
f_b(m)=\lim_{L\to\infty}\!-\frac{\kBT}{mL}\ln\CZ_{m,L}\!=\!-\kBT\left[\ln2\!+\!\left\{
\begin{array}{cc}
A_m\left(\frac{1}{2}\right)\,,& T>T_c\,,\\
\ms\ms
\max\left[A_m\left(\frac{1}{2}\right),A_m(0)\right]\,,& T<T_c\,.
\end{array}
\right.
\!\!\right]
\label{fbKH1}
\ee

\end{document}